\documentclass[runningheads]{llncs}
\usepackage[T1]{fontenc}
\usepackage{graphicx}
\usepackage{color}
\usepackage[colorlinks=true,
            linkcolor=blue,
            filecolor=blue,      
            urlcolor=blue,       
            citecolor=blue
            ]{hyperref}
\usepackage{multirow}
\usepackage{amsmath}

\begin{document}
\title{Optimizing Synthetic Data for Enhanced Pancreatic Tumor Segmentation}
%
%
\author{Linkai Peng\inst{1} \and 
Zheyuan Zhang\inst{1}\and
Gorkem Durak\inst{1} \and
Frank H. Miller\inst{1} \and
Alpay Medetalibeyoglu\inst{1} \and
Michael B. Wallace\inst{2} \and
Ulas Bagci\inst{1}}
\authorrunning{L. Peng et al.}
%
\institute{Department of Radiology, Northwestern University, Chicago, IL, USA \and
Division of Gastroenterology and Hematology, Mayo Clinic, Jacksonville, FL, USA}
\maketitle              

\footnote{L. Peng and Z. Zhang contributed equally. This work is supported by the NIH funding: R01-CA246704, R01-CA240639, U01-DK127384-02S1, and U01-CA268808.}
\begin{abstract}
\vspace{-0.8cm}
Pancreatic cancer remains one of the leading causes of cancer-related mortality worldwide. Precise segmentation of pancreatic tumors from medical images is a bottleneck for effective clinical decision-making. However, achieving a high accuracy is often limited by the small size and availability of real patient data for training deep learning models. Recent approaches have employed synthetic data generation to augment training datasets. While promising, these methods may not yet meet the performance benchmarks required for real-world clinical use. This study critically evaluates the limitations of existing generative-AI based frameworks for pancreatic tumor segmentation. We conduct a series of experiments to investigate the impact of synthetic \textit{tumor size} and \textit{boundary definition} precision on model performance. Our findings demonstrate that: (1) strategically selecting a combination of synthetic tumor sizes is crucial for optimal segmentation outcomes, and (2) generating synthetic tumors with precise boundaries significantly improves model accuracy. These insights highlight the importance of utilizing refined synthetic data augmentation for enhancing the clinical utility of segmentation models in pancreatic cancer decision making including diagnosis, prognosis, and treatment plans. Our code will be available at \url{https://github.com/lkpengcs/SynTumorAnalyzer}.


\end{abstract}

\section{Introduction}
Pancreatic cancer, the most dangerous tumor type, leads to merely a 12\% five-year survival rate regardless of the stage, according to the report from the American Cancer Society's 2023 Cancer Facts~\cite{islami2024american,siegel2023cancer}. Particularly, the survival rate drops to 3\% in distant (stage IV or metastatic) pancreatic tumors. Thus, it is vital for clinicians to detect pancreatic cancer at early stages. Computed Tomography (CT) imaging methods serve as the fundamental tool for early detection due to their non-invasive and low-cost properties~\cite{zhang2023deep}. However, the low contrast in medical images creates a unique challenge in finding tumor regions in CT scans. Effective pancreas tumor segmentation is crucial for treatment planning, as it provides essential clinical information, including precise tumor volumes. Since the deep learning age emerged, many automatic pancreas tumor segmentation methods have been proposed~\cite{ronneberger2015u,zhang2023deep,isensee2021nnu,hatamizadeh2021swin,zhang2022dynamic}. Medical segmentation datasets, including CT and MRI modalities, are also publicly available for the pancreas~\cite{antonelli2022medical,zhang2024large,roth2016data}. However, pancreas scans with tumor cases are still limited due to data privacy concerns~\cite{antonelli2022medical}. Thus, synthetic tumor for pancreatic cancer attracts significant attention. 

Several deep learning-based approaches have been proposed for synthesizing diseased regions in various organs. For instance, Shen et al. and Lyu et al.  focused on lung tumor generation~\cite{shen2023image,lyu2022pseudo},  Shang et al. addressed the fundus~\cite{shang2023synfundus}, Billot et al. proposed methods for brain lesion synthesis brain~\cite{billot2023synthseg}. Similarly, liver and other organs are studied by~\cite{lyu2022learning,jin2021free,hu2023label}. Focusing on pancreatic tumors, Lai et al.~\cite{lai2024pixel} introduced a rule-based approach that simulates tumor growth, invasion, and death. The method assigns states to each pixel within the pancreas and evolves them based on predefined rules, generating tumors at various stages.  Li et al.~\cite{li2023early} took a different approach. Authors first select locations within the pancreas and generate textures, subsequently refining shapes through morphological operations. Statistical analysis during shape generation allows for better control and more realistic rendering of both tumor size and shape. Chen et al.~\cite{chen2024towards} employed an autoencoder model combined with a latent diffusion model~\cite{rombach2022high} to synthesize realistic tumors with similar location selection process and shape generation by~\cite{hu2023label}. Wu et al.~\cite{wu2024freetumor} further advanced this method by replacing the diffusion model with a generative adversarial network (GAN)~\cite{goodfellow2020generative}. The proposed method incorporated adversarial training through a pre-trained segmentor to enhance the reliability of the synthesized tumors.

\textbf{What do we propose?} Leveraging the current tumor generation tools, we carefully investigate the limitations inherent in generation-based frameworks and present a series of experiments designed to test specific hypotheses. In particular, we investigate the impacts of \textit{synthetic tumor size variation} and the \textit{precision of tumor boundary definitions }on tumor segmentation accuracy. Our findings indicate that (1) selecting an optimal combination of tumor sizes is crucial for achieving superior segmentation outcomes, and (2) precise tumor boundary annotations significantly enhance model performance.

\begin{figure}[h]
	\centering
	\centerline{\includegraphics[width=\textwidth]{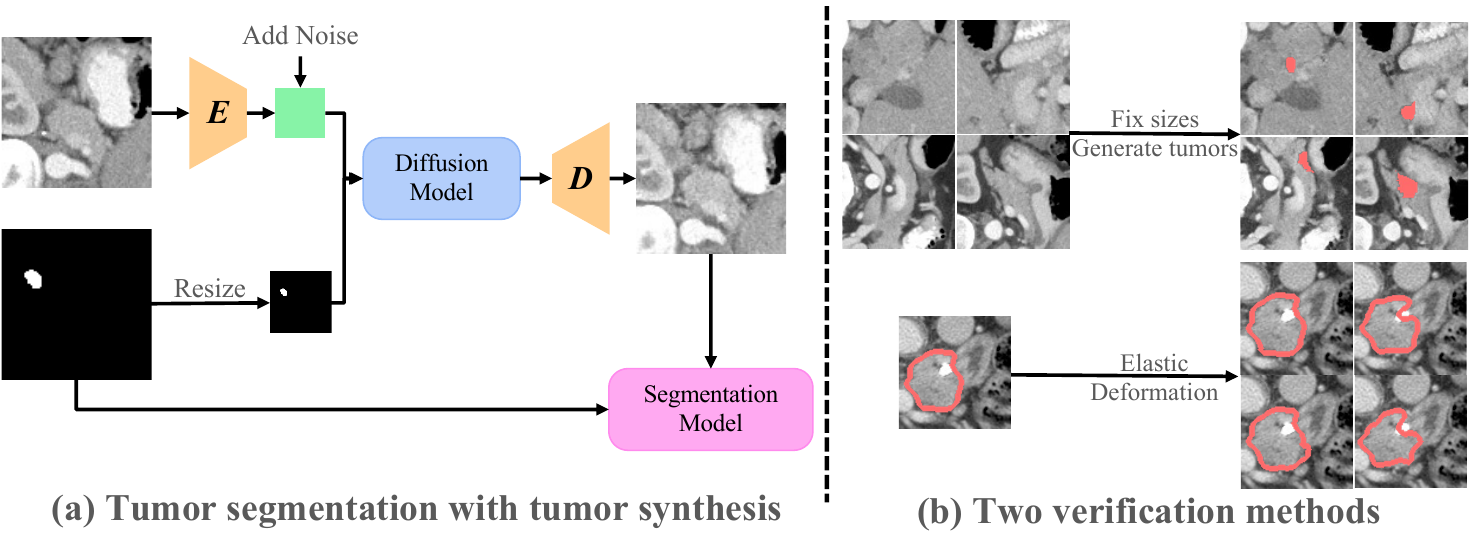}}
	\caption{Schematic demonstration of our proposed verification strategy. Panel $\textbf{(a)}$ shows the tumor segmentation process using a diffusion model to synthesize pancreatic tumors. $\boldsymbol{E}$ and $\boldsymbol{D}$ denote the encoder and decoder of a pre-trained autoencoder. Panel $\textbf{(b)}$ depicts two proposed verification methods. The upper part shows the generation of fixed-size tumors for segmentation. The lower part illustrates the elastic deformation used for generating noisy labels.}
	\vspace{-0.2cm}
	\label{main_fig}
\end{figure}

\section{Methods}
\textbf{Problem definition.}
We explore potential reasons for the unsatisfactory performance of pancreatic tumor segmentation models with tumor synthesis under real-world scenarios. Based on this analysis, we carefully evaluate the influence of tumor size and the non-perfect tumor annotations.

\textbf{Influence of tumor size.}
The size of synthetic tumors significantly influences the training of segmentation models. This factor becomes more critical when dealing with pancreatic tumors. These tumors are relatively small compared to the entire pancreas, which presents a considerable challenge for segmentation. From our observations, current methods all generate the tumor regions first and then generate the tumor textures. Given that the synthesis of the tumor texture also relies on the initial generation of the tumor region(s), the small size of these pancreatic tumors further compounds the complexity of the segmentation process.

\textbf{Influence of tumor annotations.}
Defining the boundary of synthetic tumors is another critical factor that influences segmentation models. The precision of boundary definitions directly affects a model's ability to delineate tumor margins accurately, essential for clinical applications. Since current models solely focus on the regions within the generated tumor masks, the boundaries may suffer from abrupt changes and unrealistic appearances, leading to inferior segmentation performances.

\textbf{Hypotheses:}
Based on our observations, we propose two hypotheses as follows:
\begin{itemize}
    \item \textbf{Hypothesis 1}: Effectively utilizing different synthetic tumor sizes can help improve the segmentation performances.
    \item \textbf{Hypothesis 2}: More accurate tumor boundaries can lead to better performances.
\end{itemize}

\subsection{Tumor Generation Network}
In order to carefully investigate the influence of tumor size and annotation precision, we follow the widely used framework in \textit{Difftumor}~\cite{chen2024towards} for tumor generation. An input volume \(x\) is initially processed through the encoder \(\boldsymbol{E}\) of a pre-trained autoencoder, which compresses it into latent features \(x_l\) within the latent space. Simultaneously, a corresponding tumor mask \(y\) is generated and resized to align with the dimensions of \(x_l\). These two components are integrated to form the masked features \(z = (1 - y) \otimes x_l\). Subsequently, \(z\) is fed into a pre-trained latent diffusion model~\cite{rombach2022high}. The resultant output is then upsampled through the autoencoder's decoder $\boldsymbol{D}$ to produce the final synthetic volume $\hat{x}$. $\hat{x}$ and $y$ are then used to train a segmentation model, which is supervised using both Dice loss and Cross Entropy loss. Our proposed verification methods overview is shown in Fig. \ref{main_fig}. 

\subsection{Dataset}
Following the methodology outlined in~\cite{chen2024towards}, we utilize the MSD-Pancreas~\cite{antonelli2022medical} dataset for real tumor data, while employing samples from the Pancreas-CT~\cite{roth2016data} and BTCV~\cite{landman2015miccai} datasets as healthy controls. The MSD-Pancreas dataset comprises of 282 volumetric (3D) CT scans with publicly available pixel-level annotations for pancreas and tumor regions. In contrast, the Pancreas-CT dataset includes 82 abdominal contrast-enhanced volumetric CT scans provided by the NIH team, and the BTCV dataset contains 30 volumetric CT scans with manually annotated abdominal organs including pancreas.

\subsection{Impact of tumor size}
In the original tumor generation pipeline, four pre-defined radii---\(r_{tiny}\), \(r_{small}\), \(r_{medium}\), and \(r_{large}\)---are specified. One of these radii is randomly chosen as \(r_{selected}\). Subsequently, the radii along the x, y, and z axes are generated based on a uniform distribution as shown in Eq. \ref{eq1}
\begin{equation}
    r_x, r_y, r_z \sim \text{Uniform}(r_{selected} - \Delta, r_{selected} + \Delta),
\label{eq1}
\end{equation}
where \( \Delta \) defines the variability range around the selected radius \( r_{selected} \), and is different for different tumor sizes. With this way, the model can be fed with tumors of various sizes. 

To verify our hypothesis 1, we control the tumor size during the generation process. Initially, we analyze tumor volumes from the MSD-Pancreas dataset~\cite{antonelli2022medical}, arranging them in ascending order of volume. Quartile values are then computed to categorize the tumors into four predefined size classes: tiny, small, medium, and large. Associated radii---\( r_{tiny} \), \( r_{small} \), \( r_{medium} \), and \( r_{large} \)---are determined for each category. These quartile-based radii replace the initially predefined radii to ensure an accurate representation of natural tumor size variations. This methodology facilitates the controlled synthesis of tumors, thus allowing for a systematic evaluation of the performances of segmentation models across a spectrum of tumor sizes.

\subsection{Impact of boundary}
In the original framework, tumor boundaries are refined through elastic deformation, introducing a randomized, non-linear transformation to the synthetic tumor regions. This is achieved by manipulating a grid of control points, where each control point is subjected to random shifts governed by a normal distribution with a specified standard deviation \( \sigma \). The shifts generate a displacement field \(\mathbf{D}\), described mathematically as Eq. \ref{eq2}
\begin{equation}
    \mathbf{D}(i, j) = (\Delta x, \Delta y) \sim \mathcal{N}(0, \sigma^2),
\label{eq2}
\end{equation}
where \((i, j)\) are the grid coordinates, and \((\Delta x, \Delta y)\) represents the displacement at each point, drawn from a normal distribution with mean 0 and variance \(\sigma^2\). This displacement field dictates the relocation of pixels within the original tumor region, thereby creating more realistic boundaries that closely mimic those of actual tumors.

To validate our second hypothesis, we apply the same elastic deformation process to the original tumor labels in the MSD-Pancreas dataset~\cite{antonelli2022medical}, resulting in (artificially) noisy labels. Given that the labels are three-dimensional, we apply elastic deformation to x, y, and z dimensions independently. Notably, we exclusively use real tumor images paired with the modified labels for training, omitting any synthetic volumes.

\section{Experiments and Results}
\subsection{Training protocol}
We partition the MSD-Pancreas~\cite{antonelli2022medical} dataset into training and testing subsets using a 4:1 ratio. Selected samples from the Pancreas-CT~\cite{roth2016data} dataset and the BTCV~\cite{landman2015miccai} dataset are used for synthesizing data. The intensities of all scans are clipped to the range [-175, 250] and then normalized to [0, 1]. During training, all inputs are randomly cropped to a size of 96 $\times$ 96 $\times$ 96 with an equal ratio of tumor to normal tissue regions. We also apply data augmentation techniques such as random rotation, random flip and random shift. For the autoencoder model~\cite{chen2017deep}, we adopt the Vector Quantized Generative Adversarial Networks (VQGAN)~\cite{esser2021taming} architecture to learn to map the original 3D volumes to a shared latent space. For the diffusion model~\cite{ho2020denoising}, we use the latent diffusion model~\cite{rombach2022high} structure with inputs from the compressed latent space of the autoencoder. For segmentation models, we use U-Net~\cite{ronneberger2015u}, nnU-Net~\cite{isensee2021nnu}, and SwinUNETR~\cite{hatamizadeh2021swin}. All methods are trained on an A100 GPU for 200 epochs.

\subsection{Influence of tumor size}
We trained U-Net~\cite{ronneberger2015u}, nnU-Net~\cite{isensee2021nnu}, and SwinUNETR~\cite{hatamizadeh2021swin} using synthetic tumor volumes of varying sizes: tiny, small, medium, large, and mixed, where the Mixed category aligns with the original \textit{Difftumor}~\cite{chen2024towards} method. Table \ref{tab:table1} presents the performance outcomes, highlighting both the Dice Similarity Coefficient (DSC) and Normalized Surface Distance (NSD) metrics. The results indicate a clear improvement in segmentation performance for all models when synthetic volumes are utilized. Notably, models using large synthetic tumor sizes achieved superior outcomes compared to those with mixed-size tumors, suggesting that larger synthetic tumors might be more effective in improving model accuracy. This observation underscores the impact of utilizing varied synthetic tumor sizes on segmentation performance, indicating that the optimal combination of these sizes could be crucial for achieving the best segmentation results.

\begin{table}[!t]
\centering
\caption{Segmentation performances on various tumor sizes. We report the Dice Similarity Coefficient (DSC) and Normalized Surface Distance (NSD). We can observe a clear improvement in segmentation performance for all models when synthetic volumes are utilized.}
\resizebox{0.8\textwidth}{!}{
\begin{tabular}{cccc}
\hline
Method                     & Tumor Size       & DSC(\%) & NSD(\%) \\ \hline
$\textbf{Without synthetic volumes}$ & & & \\
U-Net & - & 48.48 & 43.26 \\ \hline
nnU-Net & - & 50.50 & 47.32 \\ \hline
SwinUNETR & - & 45.88 & 39.76 \\ \hline
$\textbf{With synthetic volumes}$ & & & \\
\multirow{5}{*}{U-Net}     & Tiny   & 47.92   & 43.62   \\
                           & Small  & 50.07   & 46.45   \\
                           & Medium & 51.32   & 47.77   \\
                           & Large  & 54.84   & 49.76   \\ 
                           & Mixed & 54.49   & 49.41\\ \hline
\multirow{5}{*}{nnU-Net}   & Tiny   & 50.98   & 46.02   \\
                           & Small  & 51.45   & 46.91   \\
                           & Medium & 53.67   & 49.59  \\
                           & Large  & 55.99   & 50.89   \\
                           & Mixed & 52.10  & 47.14 \\ \hline
\multirow{5}{*}{SwinUNETR} & Tiny   & 54.77   & 49.81   \\
                           & Small  & 52.76   & 49.24   \\
                           & Medium & 55.16   & 49.29   \\
                           & Large  & 56.10   & 53.58   \\
                           & Mixed  & 55.50   & 51.03   \\ \hline
\end{tabular}}
\vspace{-0.4cm}
\label{tab:table1}
\end{table}

Qualitative visualizations of the segmentation results from all compared methods are presented in Fig. \ref{result_fig}. We provide the tumor segmentation results for U-Net~\cite{ronneberger2015u}, nnU-Net~\cite{isensee2021nnu}, and SwinUNETR~\cite{hatamizadeh2021swin}. These results demonstrate that the incorporation of synthetic volumes significantly enhances the models' ability to accurately segment tumors, as evidenced by the more precise delineation of tumor boundaries.

\subsection{Influence of noisy label}

\begin{figure}[!h]
	\centering
	\centerline{\includegraphics[width=\textwidth]{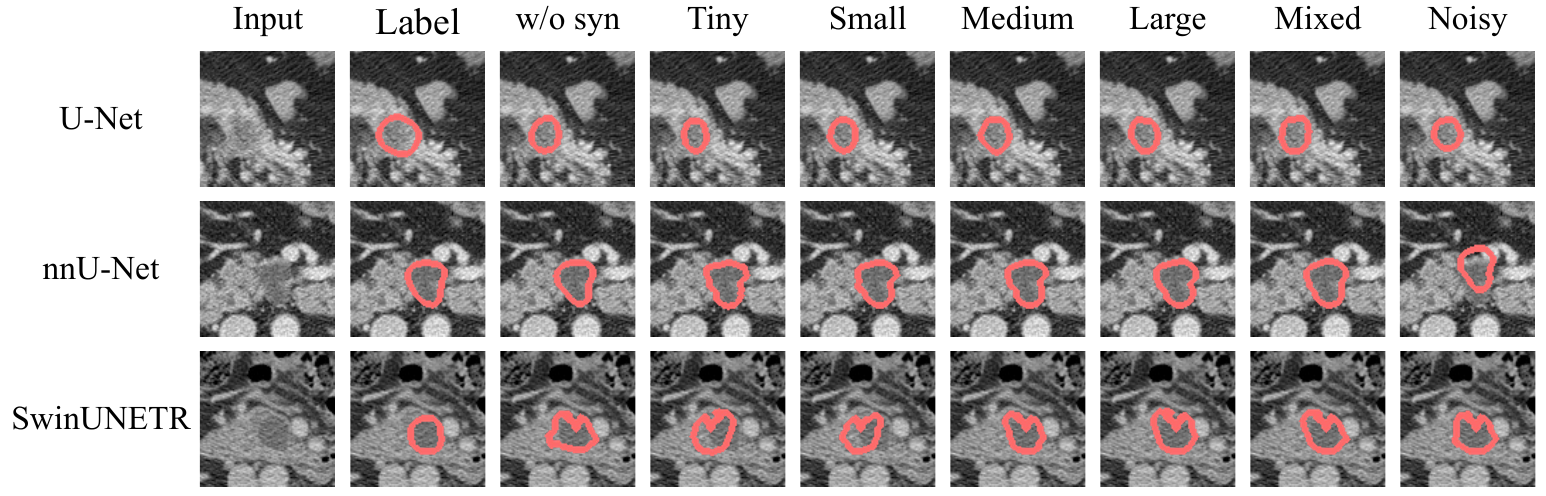}}
	\caption{Qualitative visualization results of all compared methods. The rows represent the models used, while the columns display results from left to right: raw input volumes, labels, results without synthetic volumes, and segmentation for tumors categorized as Tiny, Small, Medium, Large, Mixed, and with Noisy labels. Tumor boundaries are delineated in the figure.}\medskip
	\vspace{-0.6cm}
	\label{result_fig}
\end{figure}

We trained U-Net~\cite{ronneberger2015u}, nnU-Net~\cite{isensee2021nnu}, and SwinUNETR~\cite{hatamizadeh2021swin} using noisy labels generated by elastic deformation. Only CT scans from the MSD-Pancreas~\cite{antonelli2022medical} are used for training, and we artificially add label noise into the annotation mask. The results, detailed in Table \ref{tab:table2}, indicate a decline in performance for all models when trained with noisy labels. This can also be confirmed, as shown in Fig. \ref{result_fig}, and the segmentation accuracy declines when the models are trained with noisy labels, leading to less precise boundary delineations.

\begin{table}[!b]
\centering
\vspace{-0.2cm}
\caption{Performance comparison between original labels and noisy labels. We report the Dice Similarity Coefficient (DSC) and Normalized Surface Distance (NSD). Higher DSC and lower NSD indicate a superiority.}
\resizebox{0.6\textwidth}{!}{
\begin{tabular}{ccc}
\hline
Method      & DSC(\%) & NSD(\%) \\ \hline
$\textbf{With original labels}$ & & \\
U-Net  & 48.48 & 43.26 \\ \hline
nnU-Net  & 50.50 & 47.32 \\ \hline
SwinUNETR  & 45.88 & 39.76 \\ \hline
$\textbf{With noisy labels}$  & & \\ 
U-Net      & 43.48   & 37.90 \\ \hline
nnU-Net     & 47.41   & 41.78     \\ \hline
SwinUNETR   & 43.09   & 38.30    \\ \hline
\end{tabular}}
\label{tab:table2}
\end{table}

Additionally, we conducted an ablation study to further evaluate these models across various noise levels: low, moderate, high, and extreme. The results, presented in Table~\ref{tab:table3}, demonstrate a marked degradation in segmentation accuracy as the noise level in the labels increases. Specifically, under high and extreme noise conditions, both the Dice Similarity Coefficient (DSC) and Normalized Surface Distance (NSD) metrics for all models exhibit substantial declines. For instance, the Dice Score for U-Net, nnU-Net, and SwinUNETR drops from 43.48, 47.41, and 43.09 at the low noise level to 32.49, 37.71, and 32.14 at the extreme noise level, respectively. These results underscore the critical importance of accurate synthetic tumor boundary generation for enhancing model performance. The findings highlight that maintaining high-fidelity synthetic data is essential to mitigate the adverse effects of label noise and improve the robustness and reliability of segmentation models in clinical applications.

\vspace{-0.3cm}
\begin{table}[!htbp]
\centering
\caption{Overall performances of models with different levels of noisy labels.  We can observe a marked degradation in segmentation accuracy as the noise level in the labels increases.}
\resizebox{0.7\textwidth}{!}{
\begin{tabular}{cccc}
\hline
Method                     & Noise Level       & DSC(\%) & NSD(\%)  \\ \hline
$\textbf{With original labels}$ & & & \\
U-Net & - & 48.48 & 43.26 \\ \hline
nnU-Net & - & 50.50 & 47.32 \\ \hline
SwinUNETR & - & 45.88 & 39.76 \\ \hline
$\textbf{With noisy labels}$ & & & \\
\multirow{4}{*}{U-Net}     & Low   & 43.48   & 37.90   \\
                           & Moderate  & 40.86   & 35.31   \\
                           & High & 41.12   & 34.94   \\
                           & Extreme  & 32.49   & 25.10   \\ \hline
\multirow{4}{*}{nnU-Net}   & Low   & 47.41   & 41.78   \\
                           & Moderate  & 50.64   & 46.05   \\
                           & High & 44.38   & 38.81   \\
                           & Extreme  & 37.71   & 32.55   \\ \hline
\multirow{4}{*}{SwinUNETR} & Low   & 43.09   & 38.30   \\
                           & Moderate  & 43.69   & 38.54   \\
                           & High & 40.94   & 34.80  \\
                           & Extreme  & 32.14   & 26.42   \\ \hline
\end{tabular}}
\label{tab:table3}
\vspace{-0.8cm}
\end{table}

\section{Discussion and Conclusion}
This work investigates the performance of leading deep learning segmentation models for pancreatic tumors. We leverage the \textit{Difftumor} framework~\cite{chen2024towards} to generate synthetic tumors and evaluate three established models: U-Net~\cite{ronneberger2015u}, nnU-Net~\cite{isensee2021nnu}, and SwinUNETR~\cite{hatamizadeh2021swin}. We hypothesize that incorporating synthetic tumors and refining their properties can improve segmentation accuracy. Our comprehensive experiments demonstrate that augmenting training data with synthetic tumors significantly enhances the models' ability to delineate tumor boundaries. Notably, the size of these synthetic tumors plays a critical role in segmentation performance. Furthermore, we investigate the impact of label noise on synthetic tumor boundaries. We demonstrate that accurate tumor boundary generation is another key to better segmentation performances



This study emphasizes the critical role of high-fidelity and well-controlled synthetic data for achieving superior segmentation results in pancreatic tumors. Our findings suggest that future research should focus on developing more sophisticated methods for generating synthetic data that closely resembles real-world pathological presentations. By achieving this level of realism, we can train models with enhanced segmentation capabilities, ultimately leading to improved applicability and effectiveness in clinical practice.

%
%
%
\newpage
\bibliographystyle{splncs04}
\bibliography{ref}

\begin{thebibliography}{10}
\providecommand{\url}[1]{\texttt{#1}}
\providecommand{\urlprefix}{URL }
\providecommand{\doi}[1]{https://doi.org/#1}

\bibitem{antonelli2022medical}
Antonelli, M., Reinke, A., Bakas, S., Farahani, K., Kopp-Schneider, A.,
  Landman, B.A., Litjens, G., Menze, B., Ronneberger, O., Summers, R.M.,
  et~al.: The medical segmentation decathlon. Nature communications
  \textbf{13}(1), ~4128 (2022)

\bibitem{billot2023synthseg}
Billot, B., Greve, D.N., Puonti, O., Thielscher, A., Van~Leemput, K., Fischl,
  B., Dalca, A.V., Iglesias, J.E., et~al.: Synthseg: Segmentation of brain mri
  scans of any contrast and resolution without retraining. Medical image
  analysis  \textbf{86},  102789 (2023)

\bibitem{chen2017deep}
Chen, M., Shi, X., Zhang, Y., Wu, D., Guizani, M.: Deep feature learning for
  medical image analysis with convolutional autoencoder neural network. IEEE
  Transactions on Big Data  \textbf{7}(4),  750--758 (2017)

\bibitem{chen2024towards}
Chen, Q., Chen, X., Song, H., Xiong, Z., Yuille, A., Wei, C., Zhou, Z.: Towards
  generalizable tumor synthesis. arXiv preprint arXiv:2402.19470  (2024)

\bibitem{esser2021taming}
Esser, P., Rombach, R., Ommer, B.: Taming transformers for high-resolution
  image synthesis. In: Proceedings of the IEEE/CVF conference on computer
  vision and pattern recognition. pp. 12873--12883 (2021)

\bibitem{goodfellow2020generative}
Goodfellow, I., Pouget-Abadie, J., Mirza, M., Xu, B., Warde-Farley, D., Ozair,
  S., Courville, A., Bengio, Y.: Generative adversarial networks.
  Communications of the ACM  \textbf{63}(11),  139--144 (2020)

\bibitem{hatamizadeh2021swin}
Hatamizadeh, A., Nath, V., Tang, Y., Yang, D., Roth, H.R., Xu, D.: Swin unetr:
  Swin transformers for semantic segmentation of brain tumors in mri images.
  In: International MICCAI Brainlesion Workshop. pp. 272--284. Springer (2021)

\bibitem{ho2020denoising}
Ho, J., Jain, A., Abbeel, P.: Denoising diffusion probabilistic models.
  Advances in neural information processing systems  \textbf{33},  6840--6851
  (2020)

\bibitem{hu2023label}
Hu, Q., Chen, Y., Xiao, J., Sun, S., Chen, J., Yuille, A.L., Zhou, Z.:
  Label-free liver tumor segmentation. In: Proceedings of the IEEE/CVF
  Conference on Computer Vision and Pattern Recognition. pp. 7422--7432 (2023)

\bibitem{isensee2021nnu}
Isensee, F., Jaeger, P.F., Kohl, S.A., Petersen, J., Maier-Hein, K.H.: nnu-net:
  a self-configuring method for deep learning-based biomedical image
  segmentation. Nature methods  \textbf{18}(2),  203--211 (2021)

\bibitem{islami2024american}
Islami, F., Baeker~Bispo, J., Lee, H., Wiese, D., Yabroff, K.R., Bandi, P.,
  Sloan, K., Patel, A.V., Daniels, E.C., Kamal, A.H., et~al.: American cancer
  society’s report on the status of cancer disparities in the united states,
  2023. CA: A Cancer Journal for Clinicians  \textbf{74}(2),  136--166 (2024)

\bibitem{jin2021free}
Jin, Q., Cui, H., Sun, C., Meng, Z., Su, R.: Free-form tumor synthesis in
  computed tomography images via richer generative adversarial network.
  Knowledge-Based Systems  \textbf{218},  106753 (2021)

\bibitem{lai2024pixel}
Lai, Y., Chen, X., Wang, A., Yuille, A., Zhou, Z.: From pixel to cancer:
  Cellular automata in computed tomography. arXiv preprint arXiv:2403.06459
  (2024)

\bibitem{landman2015miccai}
Landman, B., Xu, Z., Igelsias, J., Styner, M., Langerak, T., Klein, A.: Miccai
  multi-atlas labeling beyond the cranial vault--workshop and challenge. In:
  Proc. MICCAI Multi-Atlas Labeling Beyond Cranial Vault—Workshop Challenge.
  vol.~5, p.~12 (2015)

\bibitem{li2023early}
Li, B., Chou, Y.C., Sun, S., Qiao, H., Yuille, A., Zhou, Z.: Early detection
  and localization of pancreatic cancer by label-free tumor synthesis. arXiv
  preprint arXiv:2308.03008  (2023)

\bibitem{lyu2022pseudo}
Lyu, F., Ye, M., Carlsen, J.F., Erleben, K., Darkner, S., Yuen, P.C.:
  Pseudo-label guided image synthesis for semi-supervised covid-19 pneumonia
  infection segmentation. IEEE Transactions on Medical Imaging  \textbf{42}(3),
   797--809 (2022)

\bibitem{lyu2022learning}
Lyu, F., Ye, M., Ma, A.J., Yip, T.C.F., Wong, G.L.H., Yuen, P.C.: Learning from
  synthetic ct images via test-time training for liver tumor segmentation. IEEE
  transactions on medical imaging  \textbf{41}(9),  2510--2520 (2022)

\bibitem{rombach2022high}
Rombach, R., Blattmann, A., Lorenz, D., Esser, P., Ommer, B.: High-resolution
  image synthesis with latent diffusion models. In: Proceedings of the IEEE/CVF
  conference on computer vision and pattern recognition. pp. 10684--10695
  (2022)

\bibitem{ronneberger2015u}
Ronneberger, O., Fischer, P., Brox, T.: U-net: Convolutional networks for
  biomedical image segmentation. In: Medical image computing and
  computer-assisted intervention--MICCAI 2015: 18th international conference,
  Munich, Germany, October 5-9, 2015, proceedings, part III 18. pp. 234--241.
  Springer (2015)

\bibitem{roth2016data}
Roth, H.R., Farag, A., Turkbey, E., Lu, L., Liu, J., Summers, R.M.: Data from
  pancreas-ct. the cancer imaging archive. IEEE Transactions on Image
  Processing  \textbf{10}, ~K9 (2016)

\bibitem{shang2023synfundus}
Shang, F., Fu, J., Yang, Y., Huang, H., Liu, J., Ma, L.: Synfundus: A synthetic
  fundus images dataset with millions of samples and multi-disease annotations.
  arXiv preprint arXiv:2312.00377  (2023)

\bibitem{shen2023image}
Shen, Z., Ouyang, X., Xiao, B., Cheng, J.Z., Shen, D., Wang, Q.: Image
  synthesis with disentangled attributes for chest x-ray nodule augmentation
  and detection. Medical Image Analysis  \textbf{84},  102708 (2023)

\bibitem{siegel2023cancer}
Siegel, R.L., Miller, K.D., Wagle, N.S., Jemal, A.: Cancer statistics, 2023.
  CA: a cancer journal for clinicians  \textbf{73}(1) (2023)

\bibitem{wu2024freetumor}
Wu, L., Zhuang, J., Ni, X., Chen, H.: Freetumor: Advance tumor segmentation via
  large-scale tumor synthesis. arXiv preprint arXiv:2406.01264  (2024)

\bibitem{zhang2022dynamic}
Zhang, Z., Bagci, U.: Dynamic linear transformer for 3d biomedical image
  segmentation. In: International Workshop on Machine Learning in Medical
  Imaging. pp. 171--180. Springer (2022)

\bibitem{zhang2024large}
Zhang, Z., Keles, E., Durak, G., Taktak, Y., Susladkar, O., Gorade, V., Jha,
  D., Ormeci, A.C., Medetalibeyoglu, A., Yao, L., et~al.: Large-scale
  multi-center ct and mri segmentation of pancreas with deep learning. arXiv
  preprint arXiv:2405.12367  (2024)

\bibitem{zhang2023deep}
Zhang, Z., Yao, L., Keles, E., Velichko, Y., Bagci, U.: Deep learning
  algorithms for pancreas segmentation from radiology scans: A review. Advances
  in Clinical Radiology  \textbf{5}(1),  31--52 (2023)

\end{thebibliography}

\end{document}